\documentclass[11pt, oneside]{article}   	
\usepackage{geometry}                		
\usepackage{graphicx}				
\usepackage{amssymb}
\usepackage{caption}
\usepackage{subcaption}
\usepackage{authblk}
\usepackage{booktabs}
\usepackage{tabularx}
\usepackage{rotating}

\title{Maximizing Diver Score by Examining Discrepancies in Diver Competency and Judges' Marks}
\author{Monnie McGee\thanks{mmcgee@smu.edu}}
\affil{Department of Statistics and Data Science\\ Southern Methodist University\\ Dallas, TX USA}
\date{January 4, 2025}							

\begin{document}

\maketitle
\section*{Abstract}
Central to diving competitions is the diver's ``dive list'', which is the list of dives an athlete will perform during a competition. Creating a dive list that contains enough difficulty to be competitive yet not beyond the capability of the diver is an important consideration in diving. In this work, we examine the discrepancy between a diver's ability and judges' scores in springboard diving meets with the purpose of discovering biases in scoring that might aid a diver in completing a dive list. As a measure of the ability of a diver, we calculate a mean score for all dives and all meets in which the diver has participated. We call this mean score a diver's competency score. We use the difference between judges' scores within a given meet and the diver's competency to define a discrepancy: the difference between a judge's estimation of a diver's ability and their true ability. The notions of competency and discrepancy are applied to a data set, gathered from \texttt{divemeets.com} for high-school one meter diving competitions in the US from 2017 to 2022.
\vspace{5mm}

\noindent{\bf Key Words:} scoring bias, subjective scores, athlete ability, discrepancy, springboard diving competitions

\section*{Introduction}
Research on judging bias in subjectively scored sports has gained importance due to increased media scrutiny and financial stakes \cite{sandro}. Studies have examined various aspects of judge performance, including nationalistic bias in diving \cite{emersontwo}, scoring patterns in pairs ice skating \cite{looney2004}, and the relationship between judges' scores and objective measurements in diving \cite{luedeker2022}.  In artistic gymnastics, a marking score for judges that takes into account the true ability of a gymnastics has been developed as a way to provide constructive feedback to judges and assign the best judges to the most important competitions \cite{sandro}. Video analysis has shown promise in supplementing judges' scores with objective measurements in diving \cite{luedeker2022}, and principal components analysis has been used to mathematically model subjective scoring in competitive diving \cite{young}. These studies highlight the importance of monitoring judge performance and developing methods to enhance the fairness and accuracy of scoring in subjectively judged sports.

Springboard diving presents significant technical challenges due to the complex interplay of biomechanical factors. Divers must balance various factors, including board clearance, joint constraints, and the need for in-flight adjustments \cite{miller2001}. The balance of these factors requires significant strength and agility. When done well, diving is beautiful and awe-inspiring to watch.  Diving, swimming, synchronized diving, and water polo, are often lumped together as ``aquatics'', and together they are among the top 3 sports searched on the Internet during the Olympics, and have the second most newspaper articles published during the Olympics \cite{topend}.  Diving is also a popular high school sport and collegiate sport, and aquatics as a whole is ranked as the 5th easiest scholarship to get for both men and women, just behind football and baseball for men and soccer and field hockey for women \cite{kruhm2021}. Generally, 85\% of the 14 total D1 scholarships are allocated to the swim team and 15\% to the dive team \cite{ncsa}. In the 2023-24 school year, the latest year for which numbers are available, swimming and diving is the 9th most popular high school sport for girls, with 138,174 participating, and the 10th most popular for boys, with 116,799 participating \cite{nfhsSurvey}.  For all its popularity and beauty, aquatics performance in general and diving performance in particular are not readily analyzed.

Some of the lack data analysis for the sport of diving might be due to a perceived lack of access to diving scores. However, there are several websites devoted to diving meets. DiveMeets.com \cite{divemeets} is one of the most comprehensive. It has scores for all dives, all judges, and all meets in the US including NCAA, high school, and youth diving meets. Registration is required to access the scores. The World Aquatics website \cite{worldaquatics} has scores broken down by round, diver, and dive, for elite diving meets around the world. USA Diving is another website that contains results on meets worldwide, including national Olympic trials \cite{usadiving}. Swimming.org \cite{swimming}, a website run by Swim England, has results from world and British diving championships, including the British National Cup from 2010- 2023 and the British Elite Junior Diving Championships from 2006 - 2023. SwimSwam \cite{swimswam} maintains a scoring archive of results for Olympic, international, NCAA DI, DII, and DIII, for swimming and diving, although the diving scores are typically not broken down by round and judges' score. Although not all scores are easily accessible, there is no shortage of places to obtain information on diving scores.

Most of the literature in diving has centered around analysis of nationalistic bias in judging scores in international competitions. There is a prominent  history of scandals in ice skating \cite{usatoday} and gymnastics \cite{si2012} to feed the imagination.  Evidence of nationalistic favoritism was discovered using data from diving competitions in the 2000 Summer Olympic Games \cite{emersonone, emersontwo}. The data considered for this paper are from high school competitions, and there is no information on the judge affiliation; therefore, it is not possible to spin juicy tales about bias for judges from rival high schools.  

While it gets the most press, nationalistic bias is not the only type of bias. World Aquatics (WA) warns against reputation bias, round bias, dive-order bias, difficulty bias, seeding bias, and conformity bias \cite{fina}. Table \ref{tab:bias} explains the different types of bias. 
\begin{table}[htbp]
   \centering
   \centering
    \begin{tabularx}{\textwidth}{lX}
    \toprule
    Type of Bias    & Description \\
    \midrule
Calibration & Adjusting the score for a current dive to ``make up'' for a previous score that does not match the rest of the judging panel.\\
Conformity     & Giving a better (worse) score to a diver who follows a diver who performed well (poorly).\\
Difficulty   &Giving a better score than deserved for dive with a high degree of difficulty.\\
Dive-order  & Awarding a diver a better (or worse) score than deserved  due to the score of that diver's previous dive. \\
Reputation       & Allowing the reputation or past performance of a diver to influence the score, either negatively or positively.   \\
Round &  Adjustment of scores related to the round of the competition that does not reflect a diver's ability.  \\
Seeding & Awarding higher (lower) scores to divers who are seeded first (last) within a particular competition, regardless of previous reputation.\\
Sequential & Giving higher scores to dives at the beginning or end of a round.  \\
    \bottomrule
    \end{tabularx}%
   \caption{Types of bias as defined by the FINA ``Psychology of Judging'' website \cite{fina}.}
   \label{tab:bias}
\end{table}

These biases have been shown to be present in diving competitions. Waguespack and Salomon (2016) found strong association between past performance and current performance in Olympic sports, particularly for those where the winner is determined on the basis of subjective scores \cite{waguespack}. Bruno (1986) used data from the 1983 USA Indoor Diving Nationals to examine the relationship between judges' scores and the degree of difficulty (DD) for a dive. He found that judges tended not to penalize difficult dives as much as expected, and suggested a revision of the DD tables \cite{bruno}. Other articles have investigated different types of bias, such as sequential bias \cite{kramer}, where the score of the current athlete under consideration is affected (positively or negatively) by the score of the previous athlete in the sequence. One paper analyzed the width, height, and size of the splash, which is often related to the judges' score \cite{driscoll}. 

Previous work has found evidence of difficulty bias in gymnastics, and furthermore that the difficulty bias is strong for female gymnasts and weak for male gymnasts. However, order bias does not seem to be present in the gymnastics data examined thus far \cite{morgan2014, rotthoff}. The data set that we have does not have information on the order of the divers within a given competition; nor do we have information on a diver's seed for a given competition. Therefore, we cannot comment on conformity bias, order bias, or seeding bias. This paper focuses on {\it discrepancy}, which is defined as a difference between the score given by a judge and the diver's ability \cite{sandro, emersonone}. This is important for athletes when creating their dive lists (explained in the next section) for a meet. If there is a systematic tendency for judges to give larger scores than expected at the beginning of a meet, for example, then the diver can plan for more difficult dives at the beginning of a meet. If there is a tendency for certain positions or directions to be judged more favorably, the diver can plan to perform dives in those positions or directions within the confines of meet rules. 

The data for this analysis are from one-meter springboard diving competitions at the high school level in the United States. Competitions for 25 of the 50 states are given in the data set, with California, North Carolina, and Georgia hosting the most meets. There are two basic types of meets in high school diving: dual meets (where two schools – although it could be more) compete against each other with one school as the host in their home natatorium. For these meets, there are one or two swimming officials to check strokes and touches for relay teams; however, the diving judges are typically the diving coaches of the participating teams and perhaps another coach. For these meets, the divers typically perform six dives. For more prestigious meets, which can take place at any time during the season, divers perform 11 dives. For these more prestigious meets, judges must have the appropriate certification. There is no mandatory nationwide certification for high school diving judges; however, each state high school athletic association may recommend or require a National Federation of High Schools (NFHS) course for judge certification, and most states require prospective judges to complete a training course on diving rules, judging criteria, and scoring systems \cite{nfhs_manual_2024}. For example, in Texas, the governing body for high school athletic and academic competitions is the University Interscholastic League (UIL). The UIL requires that diving judges for UIL sponsored meets be certified through a course administered by the Texas Interscholastic Swimming and Diving Officials (TISDO) organization (https://tisdo2.org/). Other states have similar certification organizations, such as the North Carolina High School Athletic Association, the California Interscholastic Federation, and the Georgia High School Association. The details of the certification are beyond the scope of this paper; however, diving judges for more prestigious meets are trained in rules of the sport and scoring systems and criteria for diving. 

First, we explain the intricacies of diving as a sport and the logistics of scoring a diving competition. The scoring for difficulty and execution differs from ice skating and gymnastics competitions. The next section explains the data set and the variables contained within. Next, we examine difficulty bias in terms of diver ability, where the definition of diver ability is the average score a diver has obtained over all competitions. Such a measure of athlete ability has been used previously in the context of gymnastics \cite{sandro} and diving \cite{emersonone}. Then, we examine overall difficulty bias and difficulty bias in the context of dive order, which is synonymous with the round of the competition. Finally, we conclude with a discussion and future directions for research.

\section*{High School Diving and its Scoring}

While some of the explanations below apply to all levels of diving, the scoring and meet structure to high school diving competitions. Regardless of diving level, every dive has three elements to it: direction, position, and rotation. The five directions are forward, backward, inward, reverse, and twist. Although a twist can be performed in any of the other four directions, historically it is considered as a fifth direction \cite{franklin}. The four positions are tuck, pike, straight, and free. The free position can be any position. Rotation has to do with the number of half-rotations that a diver does in one of the four positions and directions. For example, a forward 1.5 flip in the tuck position is a dive where the diver faces the pool on leaving the board, does one flip in the air in a tuck position, and enters the pool head first (thus 3 half-rotations). Dives performed in the same direction with the same number of rotations but a different body position are considered as the same dive. For example, a forward 1.5 in a tuck position and a forward 1.5 in a pike position are the same dive. However, a backward 1.5 somersault in the tuck position and a forward 1.5 somersault in the tuck position are different dives. The more prestigious high school diving competitions typically have 11 rounds, in which each round consists of a dive performed by each diver in the competition. Girls and boys participate in separate competitions but are scored by the same judging panel.

Critical to each diver is his or her dive list, which is the list of dives that a diver plans to perform in a given meet. A dive list must be created prior to a meet and signed by the diver and his or her coach before it is given to the judges. Once the list is finalized, it cannot be changed. The dives on the list must be performed in the order in which they are given on the list. However, divers cannot perform just any dives they wish.  For an 11-dive competition, every diver must perform 2 dives in each direction (forward, backward, reverse, inward, or twist) and an $11^{th}$ dive of the diver's choice \cite{franklin}. More precisely, the diver performs a series of voluntary dives and optional dives. The voluntary dives consist of one dive from each of the 5 diving directions. The six optional dive consists of one dive from each of the 5 diving directions with one direction represented twice. There is no set of standard voluntary or optional dives. In other words, a voluntary dive for one diver might be counted as an optional dive for another. Furthermore, the voluntary and optional dives can be performed in any order. The only limitations are that total DD for the voluntary dives must be $\le 9.0$ and the total DD for optional dives must be a minimum of 11.5 for girls and 12.0 for boys. In addition, all five groups must be represented in the first 8 rounds of diving and all 11 dives must be totally different dives \cite{franklin}. 

Unlike gymnastics or ice skating, where the difficulty of a routine is judged separately from the execution of the routine, the degree of difficulty of a dive (DD) is is determined by World Aquatics (WA), the international governing body for aquatic sports, previously known as F\'{e}d\'{e}ration Internationale de Natation (FINA).  The DD is fixed for a given dive, and is the same regardless of the age, gender, or level of the diver. For example, dive 103B (forward 1.5 in a pike position) is assigned a DD of 1.7, and this DD is the same whether the diver is male or female, is 13 years old or 22 years old, or is a high school diver or an Olympic diver. Regardless of competition level, dives are scored on a scale of 0 to 10 in increments of 0.5, with 0 being a failed dive, and 10 being a perfect dive. Just as with the DD, there are no adjustments to the execution score for the level, gender, or age of the diver. For example, most high school scores are in the range of 4-7, with a score of 8 being quite rare. However, most Olympic diving scores are in the range of 8-10, with scores of 6-7 quite rare. 

Diving, like gymnastics, ice skating and ski-jumping, is scored by a panel of judges; therefore, there is a subjective element to a diving score.  A panel of an odd number of judges, usually 5 although sometimes 3, 7, or 9, gives their scores independently for each diver for each round of a competition. The exception to independent scoring is a failed dive. To fail a dive and give the diver a score of 0, the entire panel of judges must agree to do so. The {\it net score} for each diver in each round is the sum of a $k\%$ trimmed mean of the sorted scores per round, where $k \approx .15$ for meets with 9 judges,  $k\approx .2$ for meets with 7 judges, or $k = .3$ for meets with 5 judges. For high school competitions, the net score for a dive is  multiplied by the degree of difficulty of the dive to obtain a score, called the {\it award}. The meet score for each diver is the sum of their awards for each round. The diver with the largest sum is the winner of the meet. Judges' scores are displayed on a large board after each round; therefore, every diver, coach, judge, and spectator knows which diver has the largest score at any given time during the meet. Note that the sum of the awards, and thus the total score of the meet, includes a multiplier for difficulty of the dive; therefore, it is to a diver's advantage to perform more difficult dives.

\section*{A Description of the DiveMeets Data Set}

The data set we examine contains approximately 38 thousand rows of information for high school divers participating in meets between 2017 and 2022. The data were scraped from \texttt{divemeets.com} and all divers are high school divers aged between 14 and 19.  All of the scores are for the one-meter springboard competition for all meets in the US taking place in that five-year period. For each diver, there is a unique diver ID and for each meet, there is a unique meet ID. We have information on gender, team membership, age at time of competition, and the name of the competition for each diver. For each dive, we have degree of difficulty of the dive (1.2 to 3.2 for these data), the round in which the dive was performed (1 to 11), the dive direction, position, and number of rotations, the individual score for each judge and the aggregate score for each round (award). From these data, we calculated the net score (the trimmed mean of the judge's scores without a multiplier for degree of difficulty). We also have the date that the meet took place, from which we extracted the year. Table \ref{tab:desc} gives information on the number of unique meets, unique divers, and maximum DD for the meets with 11 dives for each year of the data set. The number of divers is broken down by gender. The maximum DD is shown because it was the only statistic from a 5-number summary of the DD for each year that differed substantially for males and females. The minimum DD for both genders is 1.2, the first quartile was 1.6, the median is 1.7, and the upper quartile was either 2.1 or 2.2 depending on the year. The median net score for each year and gender is also shown. This median is the aggregated score for each meet and gender without the multiplier for DD.  

\begin{table}[htbp]
   \centering
\resizebox{\columnwidth}{!}{ \begin{tabular}{lccccccc} 
      \toprule
      Year    & Girls & Boys & Meets & Max DD (G) & Max DD (B) & Med Score (G) & Med Score (B) \\
      \midrule
      2017 & 128 & 83 & 17 & 2.8 & 3.1 & 169.0 & 177.75 \\
      2018 & 386 & 322 & 58 & 3.1 & 3.2 & 182.5 & 175.0\\
      2019 & 360 & 324 & 59 &2.8 & 3.2 & 181.0 & 170.0 \\
      2020 & 178 & 177 & 28 & 3.0 & 3.2 & 171.0 & 164.0 \\
      2021 & 167 & 117 & 19 & 2.8 & 3.2  & 184.0 & 179.5 \\
      2022 & 183 & 178 & 34 & 2.8 & 3.2 & 182.5 & 165.5\\
      \bottomrule
   \end{tabular}}
   \caption{Numbers of unique competitors by gender, number of meets, and a summary of DD by gender and year.}
   \label{tab:desc}
\end{table}

Some divers are counted twice in Table \ref{tab:desc} because they participated in multiple years. There are a total of 1872 unique divers in the data set: 1301 participated only 1 year, 439 participated in 2 years, 108 participated in 3 years, 20 participated in 4 years, and 4 participated in 5 years. From this table, more females than males participate in diving in each year, and the maximum DD for females tends to be less than the maximum DD for males. Further, it seems that the median judges' score is larger for girls than for boys. However, this score is the net score, which does not include the DD multiplier and it is aggregated over all dives. We need a more nuanced analysis to determine whether there is a gender bias in judging in either direction. 

\section*{Measuring Diver Ability}

We now turn to an examination of the difference in a diver's ability versus their scores. We call this difference the ``discrepancy''. Discrepancy is defined in terms of a judge's mark that does not reflect a diver's ability. In other words, the definition of statistical bias applies, as given by $E(\theta - \hat{\theta})$, where $\theta$ represents a diver's true ability and $\hat{\theta}$ represents an estimate of a diver's ability as given by a judge's score. Ideally, the expected value of the difference of the diver's ability and the estimate of that ability is 0. In order to estimate discrepancy, we need a reasonable estimate for a diver's true ability, $\theta$. 

Measuring overall athlete ability in the absence of an objective measure, such as a clock or a tape measure, is a tricky thing. Diving, especially high school diving, does not have a technical judging panel, as in World Championship ice skating or gymnastics, to assure that the athletes complete the rotations and twists that a dive demands. In high school competitions, there are no announcers examining every frame of an instant replay. There is research using videos from a diving competition to measure characteristics of dives that typically differentiate a good dive from a bad one, such as height of splash, distance of entry from the end of the board, and angle of entry \cite{luedeker2022}. That work showed that the judges' scores are closely aligned with quantitative values measured from the video evidence. In only one instance would the ordinal standing of the divers have been affected in a way that mattered by a difference between scores using only the video evidence versus scores given by the judges. In other words, trained diving judges are generally quite accurate when it comes to judging dive quality within a diving competition. The same work also showed that untrained diving judges are unreliable when it comes to qualitative scores matching quantitative measurements \cite{luedeker2022}. 

To define diver ability, we compute the mean net score (not including the multiplier for DD) over all dives and all meets in which a diver has participated. There is a precedent in several previous papers on diving, ice skating, and gymnastics, for using the median score as a measure of ability \cite{emersonone, emersontwo, sandro}; however, we use the mean score is used for this analysis because there tends to be little variability in the judge's scores per diver per round. We call this average measure per diver the ``competency'' score, $C_d$, for that diver. Mathematically, 
\begin{equation}
C_d = \frac{1}{mr}\sum_{all\ meets\ and\ rounds}x_{d,m,r}
\label{eq:comp}
\end{equation}
\noindent where $x$ is the net score for each round (the trimmed mean of the judges' marks without the multiplier for DD), $m$ is the number of meets in which a diver, denoted by $d$, participates between 2017 and 2022, which is the time span of the data. $r$ is the number of rounds in the dive meet. The data set has information on meets with different numbers of rounds, but for this analysis, we use data only from 11-round meets because these meets are more likely to have well-trained judges. Failed dives (given a score of 0) were deleted from the data set. Furthermore, meets where 4 or fewer divers participated were not considered in this analysis.

For each meet, dive, and diver, we compute a discrepancy score, $D_{dmr}$, which is the difference between the diver's estimated ability (competency) and the judges' combined net score for a particular dive and meet. %
\begin{equation}
D_{dmr} = x_{dmr} - C_d 
\end{equation}
\noindent where $C_d$ is defined in Equation \ref{eq:comp} and the net score is the trimmed mean of the scores given by judges for a particular diver (d) in a particular meet (m) during a particular round (r). 

A positive discrepancy indicates that the judges' net score is greater than the diver's ability, which indicates lenient judging. A negative discrepancy indicates that the judges' score is less than the ability of the diver, which indicates strict judging. These discrepancy scores can be seen as a measure of statistical bias in the sense that we are measuring the true diver's ability ($C_d$) with an estimate of their ability ($D_dmr$); however, due to the association of the term ``difficulty bias'' with lenient scoring of more difficult routines \cite{bruno, morgan2014, rotthoff}, we use the term ``discrepancy''. We examine the discrepancy scores for different ages, genders, positions, directions, and degrees of difficulty. Note the discrepancy scores are computed from net scores (the trimmed mean of judges' marks for a given dive, diver, and meet); therefore, they do not include a multiplier for DD. In the next section we examine discrepancy by age,  gender, round, and DD. 
 
\section*{Results for an Analysis of Discrepancy }\label{sec:results}

\begin{figure}[htb]
\centering
    \begin{subfigure}[h]{0.45\textwidth}
        \includegraphics[height=2in, width=.95\textwidth]{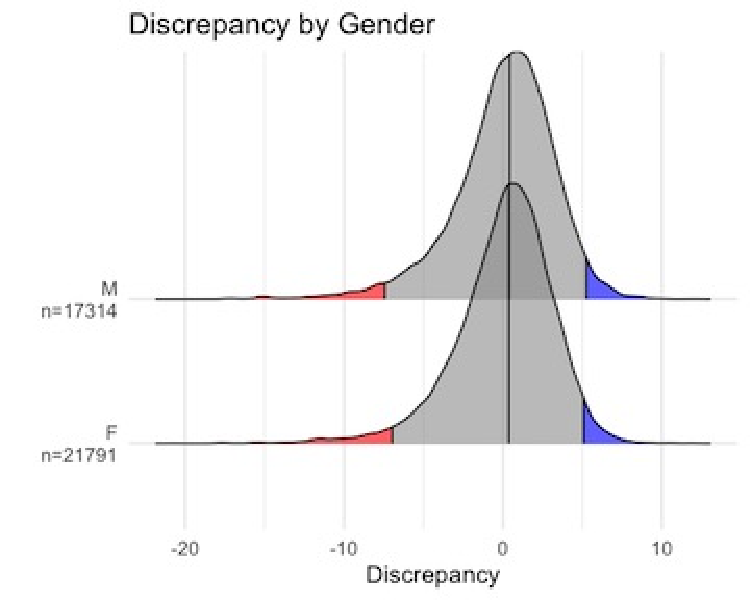}
        \caption{Discrepancy by gender. Each curve is a density estimate of the discrepancy for the gender referenced on the vertical axis.}
        \label{fig:gender}
    \end{subfigure}
        \begin{subfigure}[h]{0.45\textwidth}
        \includegraphics[height=2in,width=\textwidth]{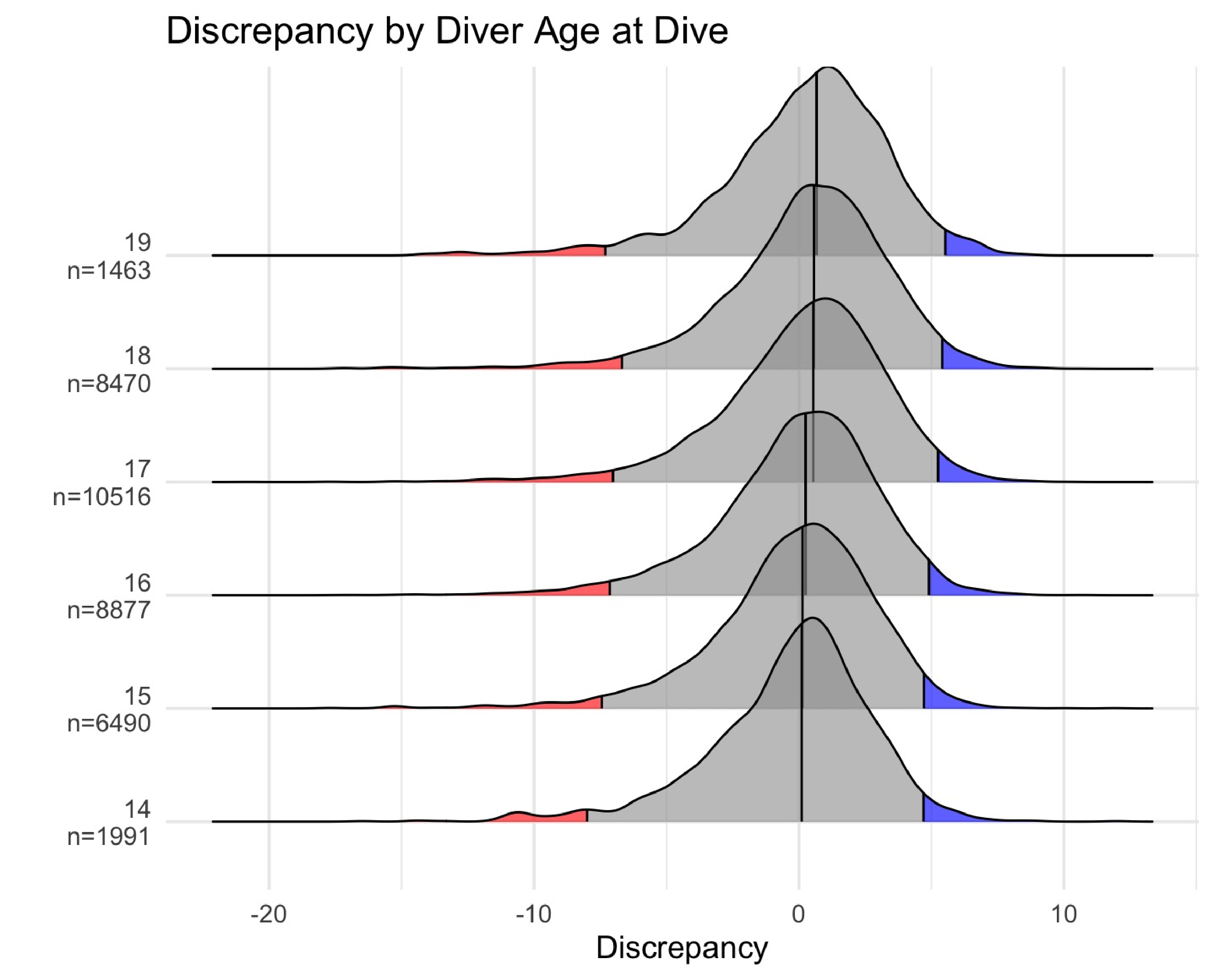}
        \caption{Discrepancy by age at dive. Each curve is a density estimate of the discrepancy for the age referenced on the vertical axis.}
        \label{fig:age}
    \end{subfigure}
\end{figure}

\begin{figure}[htb]
\centering
    \begin{subfigure}[h]{0.495\textwidth}
        \includegraphics[width=\textwidth]{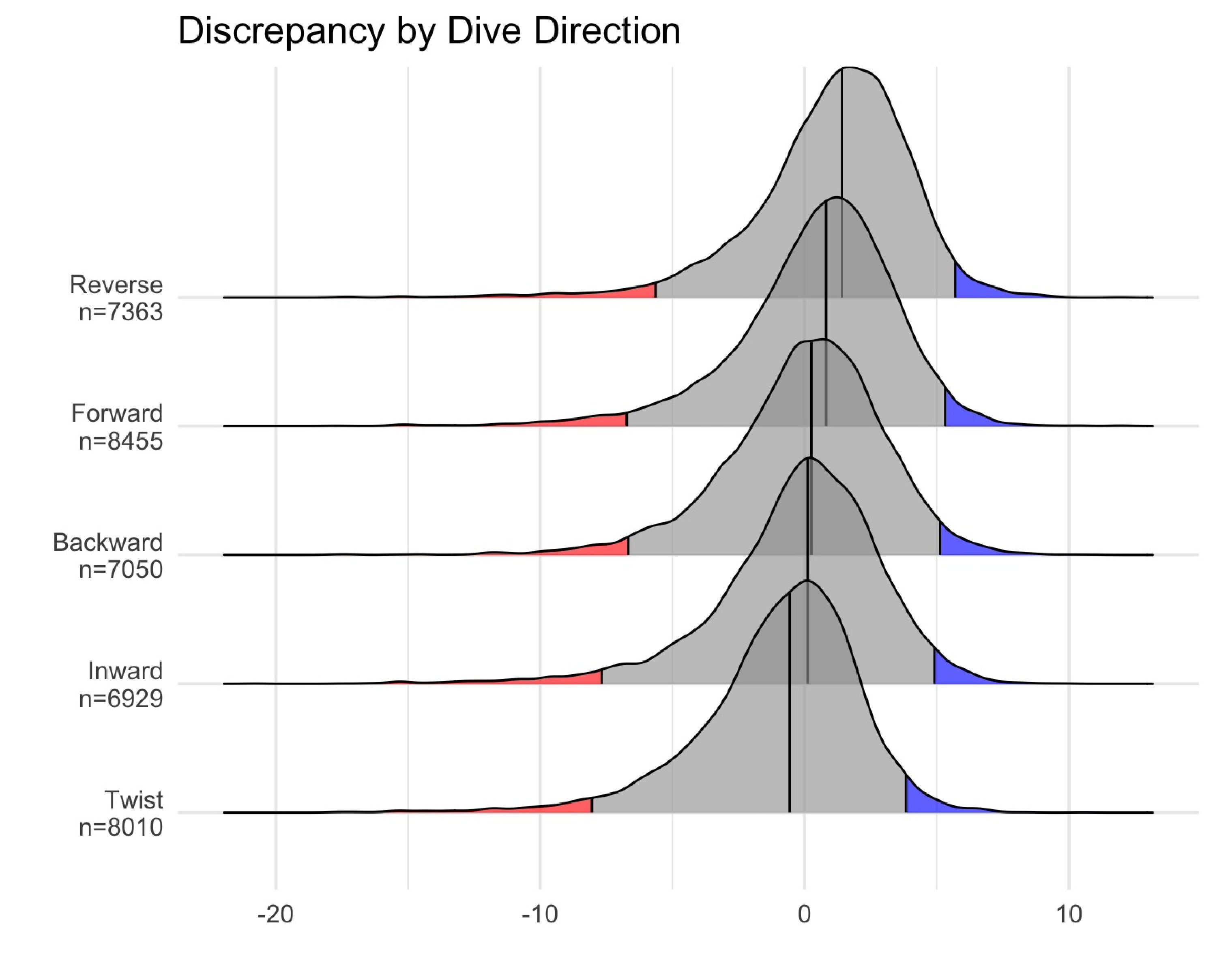}
        \caption{Discrepancy by Dive Direction}
        \label{fig:dir}
    \end{subfigure}
        \begin{subfigure}[h]{0.495\textwidth}
        \includegraphics[width=\textwidth]{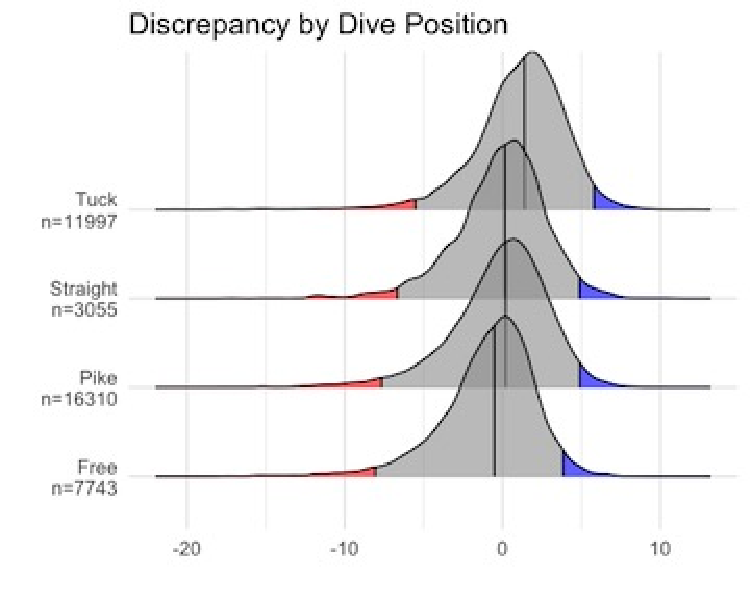}
        \caption{Discrepancy by Dive Position}
        \label{fig:pos}
    \end{subfigure}
\end{figure}

\begin{figure}
   \centering
   \includegraphics[scale=.5]{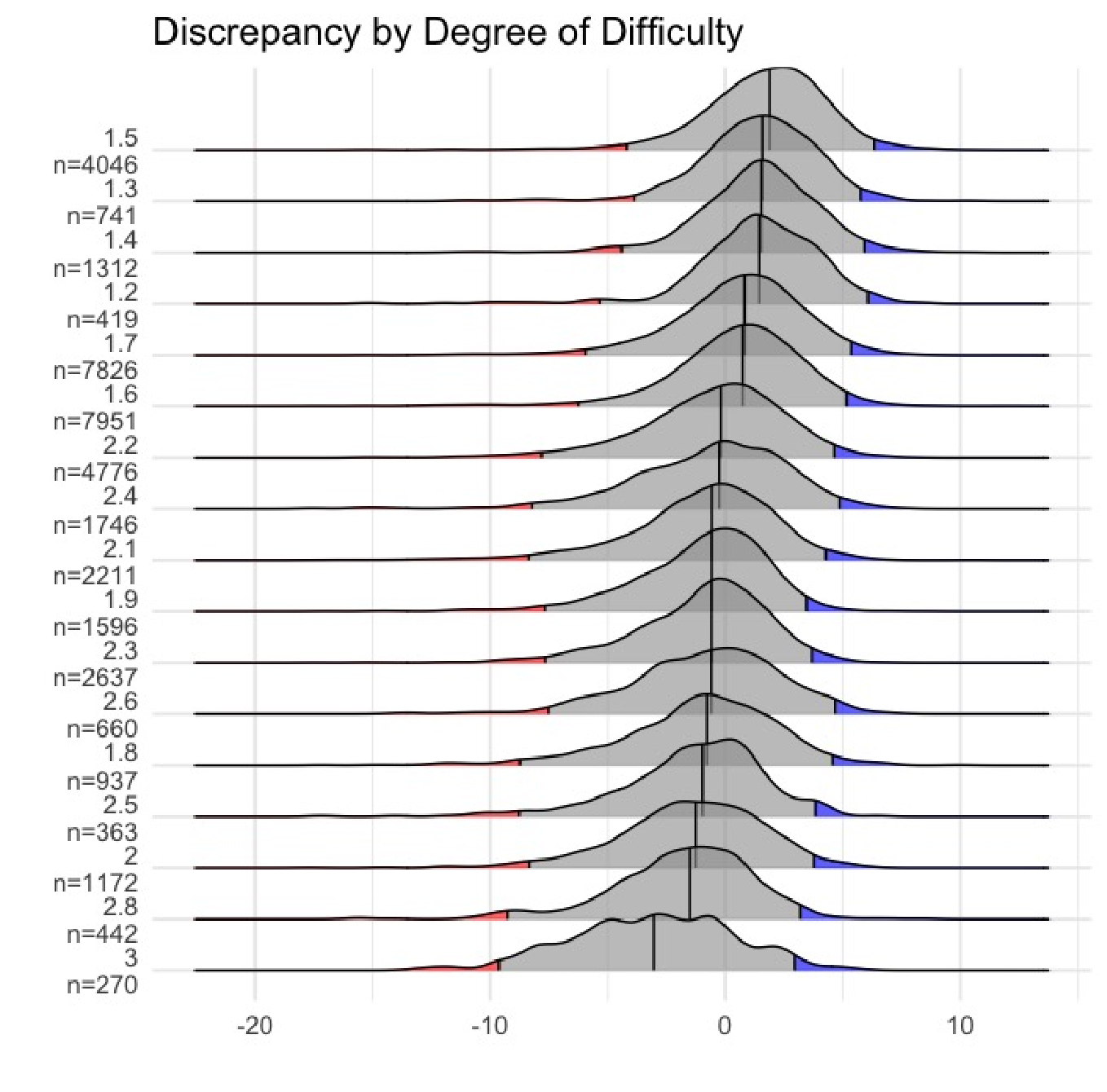} 
   \caption{Discrepancy plotted against the degree of difficulty assigned by WA for a given dive. The number of dives performed at this difficulty is on the vertical axis under each assigned DD.}
   \label{fig:discDD}
\end{figure}

For all meets, divers, and dives, ridgeline plots \cite{wilke} are used to show the distribution of the discrepancy according to gender, age, dive position, dive direction, and DD (Figures \ref{fig:gender} - \ref{fig:discDD}). The y-axis for each plot gives different categories for an explanatory variable. The x-axis is the discrepancy, which is typically in a range of -25 to 15. The black line in the middle of each curve is the median discrepancy, the red area marks the $2.5^{th}$ percentile for the discrepancies, and the blue area marks the $97.5^{th}$ percentile for the discrepancies. For all ridgeline plots, the categories are ordered by the median discrepancy. For each category, the total number of dives is given under the category name on the vertical axis.

Figure \ref{fig:gender} shows a plot of the distribution of the discrepancy for male and female divers. The black lines in the middle of each curve correspond to the median discrepancy. Visually, the black lines are aligned at 0 for both genders; numerically, the median for females is 0.36 and the median for males is 0.40, for an absolute difference of 0.04. There is an indication that judges tend to give slightly higher scores overall than is deserved; however, this effect is the same regardless of gender of the diver. Figure \ref{fig:age} shows the discrepancy plotted for each age. There is a slight shift to the right as the divers age (the medians for 14-19 year olds are 0.11, 0.14, 0.25, 0.55, 0.57, and 0.67, respectively). For 14 and 15 year old divers (who have the same median discrepancy), the scores are quite close to unbiased. A calculation of the mean and standard deviation for each age shows that both male and female divers increase their DD as they age. The change in mean DD is from 1.75 (1.80) at age 14 for female (male) divers, to 1.86 (1.91) for female (male) divers at age 19. The standard deviations for both genders also increase slightly from .2 to .4 as the age of the diver increases. An analysis for the number of years a diver has participated in diving competitions (1 to 5 for our data set) shows similar means and standard deviations of DD as  given for age of the diver. The change in median discrepancy by age might indicate improvement in diver ability with experience in diving as much as it would indicate a tendency for judges to judge older divers more leniently. 

Figures \ref{fig:dir} and \ref{fig:pos} show the distribution of discrepancy scores for dive directions and dive positions, respectively. The five directions are forward, backward, inward, reverse, and twist. To perform a forward dive, the diver faces the pool on take off and rotates forward. For a reverse dive, the diver faces the pool on take off and rotates backward toward the board (the opposite direction of a forward dive). To perform a backward dive, the diver turns his or her back to the pool and rotates backward. An inward dive is performed by turning one's back to the pool and rotating toward the board on take off. Generally, reverse and inward dives are considered more difficult than forward and backward dives in the sense that a dive in a forward direction will have a lower degree of difficulty than that same dive in a reverse or inward direction. For example, a forward 1.5 in a tuck position has a DD of 1.6. A backward 1.5 tuck has a DD of 2.0, a reverse 1.5 tuck as a DD of 2.1, and an inward 1.5 tuck has a DD of 2.2. The median discrepancy from Figure \ref{fig:dir} appears to be close to 0; however, the numerical median discrepancies are 1.41, 0.82, 0.26, 0.11, and -0.57 for the reverse, forward, backward, inward, and twist directions, respectively. Reverse dives tend to be scored more leniently, which is some indication of a ``bump'' in score for performing a more difficult dive; however, divers do not receive a similar bump for backward and inward dives, which also tend to be more difficult. Twisting dives tend to get lower scores than the diver's ability would suggest. Discrepancies for reverse, forward, and twisting dives are certainly large enough to change a diver's score substantially.

The positions are tuck, straight, pike, and free. Free is most often reserved for twisting dives, where the diver is free to twist in any position. Generally, dives in a pike or straight position will have a greater DD than the same dive in a tuck position. Numerically, the median discrepancy for the free position is -0.53 for pike it is 0.2, for straight it is 0.14 and for tuck position is it 1.39. Note that the median discrepancy for free and tuck are certainly large enough to impact a diver's final score and there is clear evidence of judges scoring dives in the tuck position more leniently.

\subsection*{Discrepancy vs. Degree of Difficulty}
For further examination of DD, the values of the DD assigned by WA were plotted on the vertical axis with the discrepancy on the horizontal axis (Figure \ref{fig:discDD}). Because few dives at this level have DD greater than 2.7, any dive with a DD between of 2.7 to 2.9 was assigned to a DD of 2.8, and any dive with DD of 3.0 or greater was assigned a DD of 3.0. Easier dives, with degrees of difficulty between 1.2 and 1.6, are more leniently judged, while more difficult dives, with DD of 2.1 or greater are more harshly judged. For example, the largest positive median score is 1.87 for dives with DD of 1.5, and the largest negative score is -3.04 for dives with DD of 3 or more. It is not strictly true that more difficult dives are more harshly judged, as dives with DD of 1.8 (median discrepancy of -.82), for example, are judged more harshly than dives with a DD or 2.3 or 2.6 (median discrepancies of -.59 and -.55, respectively). Interestingly, 86\% of the dives with DD equal to 1.8 or 1.9 are twisting dives. This explains the negative discrepancies for these purportedly easier dives, as judges tend to judge twisting dives more harshly (Figure \ref{fig:dir})

This finding would seem to be in conflict with the previous direction of difficulty bias, where more difficult dives are judged more leniently \cite{bruno, morgan2014}. However, we are not really measuring difficulty bias in the same way as was measured by previous investigators of difficulty bias in judged sports. Difficulty bias is a tendency for judges to award higher scores to more difficulty routines \cite{morgan2014}. Discrepancy is a measure of the distance of a judges' score from the diver's ability. It is possible that the expected direction of ``difficulty bias'' is reversed because a diver is attempting a dive that is beyond his or her ability; thus getting a smaller score than expected. For discrepancies in the positive direction, divers likely perform the easier dives quite well because they have been in their repertoire longer than the more difficult dives. Anecdotally, divers tend to have less practice with the more difficult dives because they have to achieve a certain level of ability before attempting these dives. In addition, more difficult dives might have not been performed in a previous competition. Therefore, the harsher judging for more difficult dives in these data is likely due to divers overreaching their ability and/or having less practice with more difficult dives in competition. There might be other factors at play, also, as we examine in the next section. 

\subsection*{Discrepancy by Round of Competition}

Another potential factor affecting discrepancy is the round within a diving meet.  Recall that each meet is 11 rounds, and, depending on the number of divers in a meet, the judges could be judging for several hours. For the analysis that follows, we have eliminated meets in which fewer than 5 divers participate in the meet.
\begin{figure}[ht]
   \centering
   \includegraphics[scale=.22]{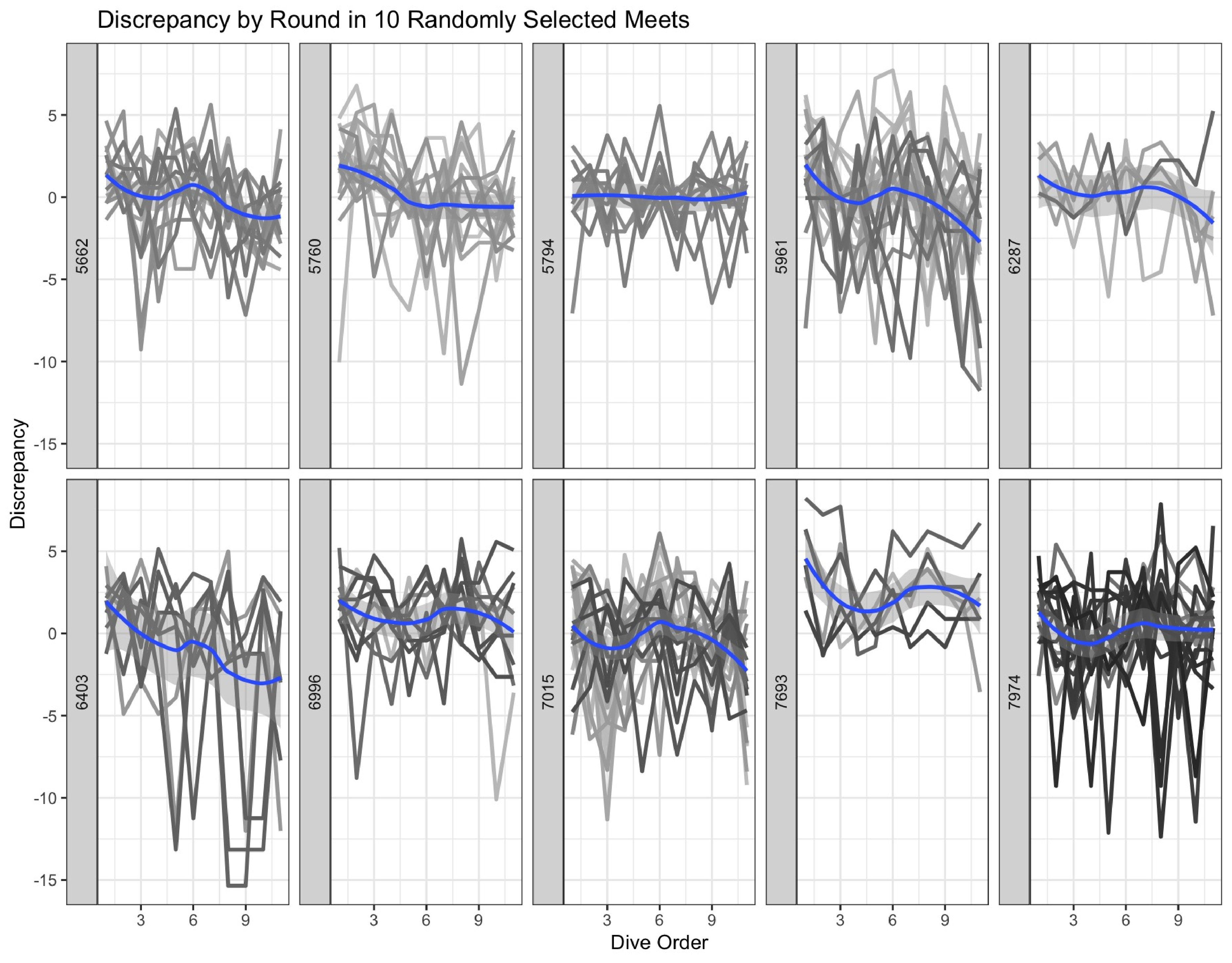} 
   \caption{Discrepancies (vertical axis) plotted as a line for each diver in 10 randomly selected meets. The horizontal axis is the round for the meet from 1 to 11. The blue line is a loess smoother. Each meet shows evidence of harsher judging (negative discrepancies) as the meet wears on.}
   \label{fig:rounds}
\end{figure}

\begin{figure}[t!]
   \centering
   \includegraphics[scale=.3]{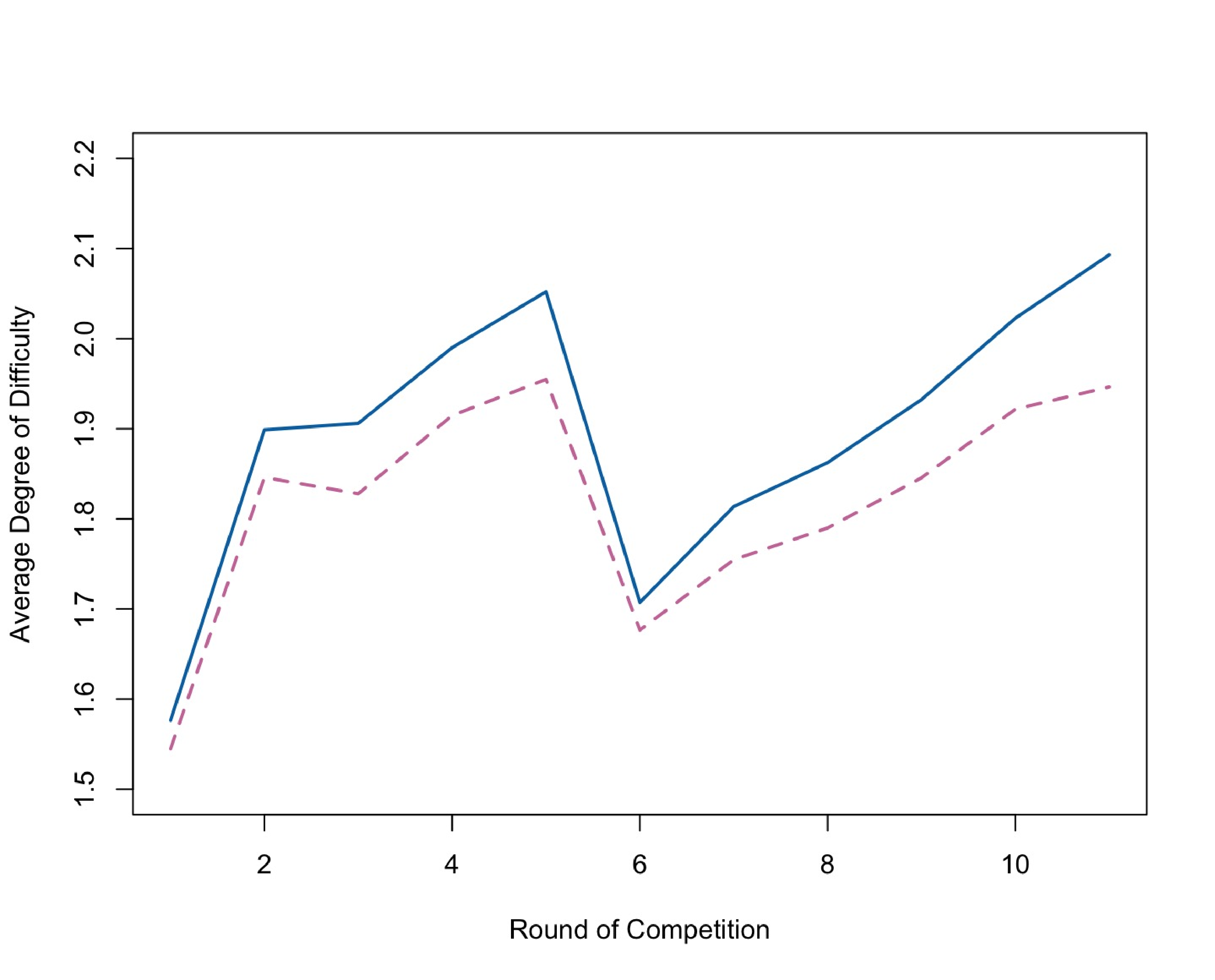} 
   \caption{DD (on the vertical axis) plotted versus the round of competition (on the horizontal axis). The magenta dashed line shows the fluctuation in DD for girls, and the blue solid line shows the fluctuation for the boys. There is a similar pattern of DD by round for each gender.}
   \label{fig:roundDD}
\end{figure}
Figure \ref{fig:rounds} shows discrepancies for all divers in ten randomly selected meets. The number of divers in the meet for this plot range from 5 to 21 (in the data set, the maximum number of divers within a meet is 45). Each frame represents a single meet, and each dark gray or black line represents the discrepancies for a different diver.  The frame label on the left hand side of each frame gives the unique meet ID. The discrepancy is plotted on the vertical axis and the round in which a dive was performed on the horizontal axes for all frames.  The blue curve within each frame is a loess smoothed line.  With the exception of meet 5794, where the loess line is flat, the loess lines indicate a trend toward a positive discrepancy at the beginning of a meet, near 0 discrepancy near the end of the meet, and a negative discrepancy at the end of the meet. Interestingly, the effect is there regardless of the meet, which is confounded with the judging panel. We also see evidence of much more variability in judging panels for some meets, as in meet 5961, 6403, and 7974. Because the data do not contain information on the identify or affiliation of the judges, we cannot determine the level of training of these judges nor their tendency to be harsh or lenient.

There are 10 randomly selected meets represented in Figure \ref{fig:rounds}, and each one of them shows various gradations of negative trends. To examine whether the negative trend in discrepancy was common throughout the data set, we calculated slopes of discrepancy versus round for each of the 215 meets in the data set. The mean slope of the discrepancy is $-0.145$ (median = $-0.147$) and the standard deviation is $0.128$. 89\% of the slopes are negative, indicating that negative discrepancies are quite common in these 11-round meets. Therefore, what is seen in Figure \ref{fig:rounds} holds true for most of the meets; there is a tendency for judges to judge more harshly as a meet enters into its final rounds.

Recall that divers are free to perform dives in almost any order for a given meet, with respect to the stipulations on voluntary vs. optional dives mentioned previously. Round and dive are confounded, and dives have different degrees of difficulty. It is possible that the negative discrepancy seen at the end of the meets is due to divers reserving more difficult dives for latter rounds (to finish with a splash, so to speak). Figure \ref{fig:roundDD} shows a line plot of the average DD for each round for girls (magenta dashed line) and boys (blue solid line). Regardless of gender, on average, divers start with less difficult dives (the most popular starting dive is a forward 1.5 in a tuck position, $DD=1.6$), work up to more difficult dives by round 5, then start again with less difficult dives and finish with more difficult dive. This overall pattern makes sense with the pattern of discrepancy in Figure \ref{fig:rounds}: divers are performing above their ability at the beginning of the competition with easier dives, taking more risks in the middle, and then finishing with a difficult dive that is perhaps beyond their ability (or perhaps they are simply tired).  The pattern of DD by round partially explains the tendency for a negative discrepancy during the final rounds of a competition.

Starting with a dive that matches a diver's ability plays into performance bias. If a diver starts with a good score, the judges might reward that diver with higher scores than expected in subsequent rounds, and performance bias might partially explain the tendency for discrepancies to be close to 0 in the middle of a meet. Dive lists typically start with a voluntary dive (the most popular starting dive is a forward 1.5 in a tuck position), which are dives which all divers should perform well, and thus provide a baseline ability and score for each diver.  The data used for this research give the round in which each dive was performed for each diver; however, the data does not designate which dives are voluntary and which are optional. It is possible to determine voluntary from optional dives in a reasonable fashion and examine the discrepancies between voluntary vs. optional dives, but this is left for future research. Other types of bias, such as sequential bias and conformity bias (see Table \ref{tab:bias}), might also be in play. Unfortunately, we do not have access to sequential diver order; that is, the data do not have an indicator for which diver went first, which went second, etc, for each meet; therefore, we cannot examine sequential bias or conformity bias with these data.  

In conclusion, visual exploration of ridgeline plots and numerical summaries for each explanatory variable indicate that round and DD are associated with largest absolute discrepancies. Most of the discrepancy values are quite small; however, recall that scores are given in increments of 0.5; therefore, a discrepancy of 0.5 or more could have a beneficial (or deleterious) effect on the overall placement of the diver within the competition.

\subsection*{Mixed Effects Model}
Figures \ref{fig:gender} to \ref{fig:rounds} show discrepancies for each of several variables as main effects. We wish to model them simultaneously in order to control for the fixed effects. We use a mixed effects model with a random intercept for diver, separated by levels of degree of difficulty. The degree of difficulty was grouped into beginner ($1.2 \le DD \le 1.5$), intermediate ($1.6 \le DD \le 2.2$), and advanced ($DD > 2.2$). For each model, Age, Gender, Position, and Direction are fixed effects. The reference for the gender factor was female. Position was entered as a factor with the straight position as the reference, and direction was entered into the model with forward as the reference. 

Table \ref{tab:mmResults} shows the parameter estimates, standard error, and t-statistics for each of the fixed effects for each grouping of DD. For beginners, the round coefficient and its test statistic are small; therefore, round does not seem to have an effect on the discrepancy for beginning dives. For intermediate and advanced dives, the coefficients on the round are small; however, the t-statistics are large. For every round that passes, the discrepancy decreases by a small amount for intermediate and advanced dives, indicating that the judges tend to judge more harshly (net scores are less than the diver's competency) as a competition continues, adjusted for age, gender, position, and direction. This finding matches the visual evidence seen in Figure \ref{fig:rounds}. The coefficient for age is 0.19 for beginning dives, .18 for intermediate dives, and 0.26 for advanced dives, which indicates that judges tend to judge more favorably for each year increase in age of the diver. This finding is commensurate with Figure \ref{fig:age}. These coefficients for age are small, but they are large enough to make a difference in the outcome of a competition because the net score changes during a meet while the diver ability remains constant. An increase (or decrease) of a fraction of a point per round could change ordinal standings for a given meet.
\begin{sidewaystable}
  \centering
\begin{tabular}{lccccccccc}
\hline
&\multicolumn{3}{c}{Beginning, $n=6314$, ICC = 4.7\%} & \multicolumn{3}{c}{Intermediate, $n=25621$, ICC = .3\%} & \multicolumn{3}{c}{Advanced, $n=5872$, ICC = 4.9\%} \\
\hline
Effect & Estimate & SE & t-statistic & Estimate & SE & t-statistic & Estimate & SE & t-statistic \\
\hline
Intercept & -2.11 & 0.48 & -4.36 & -2.18 & 0.26 & -8.52 & -6.75 & 1.36 &  -4.96 \\ 
  Round & -0.00 & 0.01 & -0.01 & -0.07 & 0.01 & -11.61 & -0.05 & 0.01 & -3.91 \\ 
  Age & 0.19 & 0.03 & 7.07  & 0.18 & 0.01 & 12.61  & 0.26 & 0.04 & 6.79 \\ 
  Male & -0.16 & 0.07 & -2.22 & 0.16 & 0.04 & 4.11  & 0.23 & 0.10 & 2.25 \\ 
  Free &           &         &          & 0.49 & 0.13 & 3.68 & 4.97 & 1.45 & 3.43 \\  
  Pike & 0.05 & 0.22 & 0.22 & -0.75 & 0.08 & -9.87 & 0.96 & 1.18 & 0.81 \\ 
  Tuck & 0.49 & 0.22 & 2.19 & 0.91 & 0.08  & 11.85 & 2.23 & 1.18 & 1.88 \\ 
  Backward & -0.38 & 0.10 & -3.69 & -0.57 & 0.06 & -8.90 & -0.82 & 0.14 & -5.86 \\ 
  Inward & & && -0.23 & 0.06 & -4.08 & -1.03 & 0.16 & -6.37 \\ 
  Reverse & 0.94 & 0.08 & 12.31 & 0.57 & 0.07 & 8.13 & 0.49 & 0.13 & 3.79 \\  
  Twist & & &  &-1.75 & 0.15 & -11.80 & -3.38 & 0.84 & -4.04 \\  
\hline
\end{tabular}
\caption{Coefficients for a Mixed Effects Model with a random intercept for Diver and Discrepancy as the response. Three models were calculated: one for beginning dives ($1.2\le DD \le 1.5$), one for intermediate dives ($1.6 \le DD \le 2.2$), and one for ($DD > 2.2$).}
\label{tab:mmResults}
\end{sidewaystable}

For gender, we see that there is a negative coefficient for beginning dives for males and a positive coefficient for intermediate and advanced dives. Thus, males tend to be judged more favorably as the degree of difficulty of a dive increases, and less favorably for dives with smaller DD. 
There are no coefficients for free, inward, and reverse for beginning dives, as beginning dives do not consist of these positions and directions. The reference position is straight, and we see that free, pike, and tuck have positive coefficients compared to straight, although the coefficient for pike has a small test statistic for beginning and advanced. Therefore, it seems that the pike position does not give any advantage in terms of discrepancy over the straight position. When considering intermediate dives, the pike position is judged more harshly than the straight position. The tuck position tends to have the largest advantage over the straight position. Regarding the directions, both backward, inward, and twist  positions tend to be judged more harshly than the forward position; the reverse position is judged more favorably regardless of the level of the dive.

Note the small size of the adjusted intraclass coefficients (ICC) for beginning, intermediate, and advanced dives (0.047, 0.003, and 0.049, respectively). We report the adjusted ICC because the ICC can be interpreted as the proportion of variance that is attributable to the ``grouping structure'' in the population \cite{gelman}. Here, the random intercept, which represents a different starting discrepancy for each diver, defines the ``grouping structure''. Therefore, only 4.7\% of the variability in discrepancy for beginning dives, .3\% of the variability in discrepancy for intermediate dives, and 4.9\% of the variability in discrepancy for advanced dives is accounted for by different starting discrepancies for each diver, adjusted for all other variables in the model.  In plain language, any effect of discrepancy that an individual diver contributes is completely overshadowed by the fixed effects. This is the opposite effect of a similar analysis done for track and field data, where the variation for the individual athlete greater than 75\% \cite{mcgee}.

\section*{Discussion and Conclusion}\label{sec:conc}

We based our calculations for diver ability and difficulty bias from previous work in ice skating and gymnastics \cite{sandro, morgan2014, rotthoff}. For gymnastics and ice skating, there are technical panels that calculate the difficulty of a routine, and the difficulty score can be challenged within certain limits. For diving, the DD of a dive is set by WA and it is not up for discussion. Furthermore, the DD for a dive is the same regardless of the gender, age, or level of the diver.  In diving there are 11 rounds consisting of one dive each, and no two divers will perform the same dives in the same order in the same meet. This makes comparisons among divers performing the same dive difficult. For example, it is quite possible that all divers within a meet will the same dive; however, they will do so at different points in the competition, which means that any differences could be due to conformity bias or sequential bias (see Table \ref{tab:bias}).  With the data we have, we do not have information on the order of each diver within each meet (sequential order); therefore, we cannot examine sequential bias or conformity bias. Furthermore, we do not have information on the seed for each diver within a meet or their reputation prior to each meet. We restricted our current study to examination of the differences in discrepancy between age,  gender, round, position, and direction in order to determine ways a diver optimize a dive list for a meet to maximize his or her score.

Results from the mixed model analyses show a positive relationship between discrepancy and age, tuck position (relative to straight) and reverse direction (relative to forward). These relationships are somewhat different depending on DD. For example, for intermediate dives, the coefficient on pike position is large and negative, while for beginning and difficult dives, the coefficient on pike position is small and positive. While we did not see a bias for gender of the diver when visually examining discrepancies by gender with ridgeline plots, there is some evidence from the coefficients of the mixed model in Table \ref{tab:mmResults} that girls and boys are scored differently with respect to DD, when controlling for round, position, direction, and age. The REML values are quite large ($> 29$K) for beginning and advanced dives and $> 127$K for intermediate dives; therefore, we have certainly not accounted for all variability. One issue is that dive order is confounded with round, because each diver can choose his or her own dive order within certain parameters. Finally, we did not have information on sequential order, which would be the order of divers within a competition. We had only the dive order in which a particular diver performed a particular dive. In later studies, we hope to obtain information on the order of the divers within a meet and to examine sequential bias and conformity bias. 

We employ three mixed effects models with a random intercept for each diver. For each model, ICC measures are close to 0, indicating little evidence of a cluster effect based on the diver. Divers are nested within meet, and each meet has a different panel of judges; therefore we can use meet as a surrogate for a panel of judges. However, each meet has a different number of divers, as seen in Figure \ref{fig:rounds}, and not every diver participates in each meet. This creates an imbalance for the purpose of modeling, and we did not attempt to deal with the imbalance in this analysis. For beginning and intermediate dives, there are few large correlations, as defined by $r > 0.2$, among the fixed effects (1\% for beginning and none for intermediate). For the advanced dives, 16\% of the correlations among fixed effects are larger than $0.2$. We did not attempt to deal with these correlations in the structure of the mixed model. Like all observational data, confounding factors exist in the diving data, and we do not have information in these data to account for several of these factors within a statistical model. 

So, what does the analysis of discrepancy indicate for a diver's dive list? The purpose of this analysis is to determine how a diver can use information about the discrepancy between judges' scores and diver ability to maximize their score in a meet. From the above analysis, we have the following advice: on average, older divers have larger positive discrepancies; therefore, stay in the sport! Male divers will benefit more from increasing the difficulty of their dives than female divers will. However, their discrepancy is slightly negative for the easier dives, which means that male divers should not increase difficulty as the expense of practicing the lower level dives. The dive list must contain inward and backward dives. Choose inward and backward dives that are in the intermediate DD category. All divers must also do twisting dives, which are typically judged more harshly; however, the tuck position and reverse direction are generally more favorable; therefore, a reverse twist in the tuck position might give a better score than the diver's ability would merit (and would look quite impressive). However, for intermediate and advanced dives, a free position is associated with a larger increase in discrepancy than is the tuck position. In general, it does not seem that the pike position increases discrepancy even though it generally increases DD for a given dive. For beginning and advanced dives, one should consider a pike position if (1) the diver is in danger of over rotating the dive in a tuck position or (2) the diver is flexible enough to perform the pike really well.  There is also a positive coefficient for the pike position for intermediate difficulty dives. And because the $11^th$ optional dive is the diver's choice, choose something other than a twist; the best bets are an intermediate level dive in a reverse direction and the pike position. 

\section*{Acknowledgements}
The author thanks Gabe Downey, University of Kanas, and Zachary Berg, Pembroke School of Kansas, for providing the data for this paper.

\bibliographystyle{plain}
\bibliography{jsm2023}

\end{document}